\documentclass[aps,twocolumn,pra,groupedaddress,superscriptaddress,amsmath,amssymb,showpacs,noeprint]{revtex4-2}

\usepackage{dcolumn}
\usepackage{bm}
\usepackage{graphicx}
\usepackage{amsmath}
\usepackage{amsfonts}
\usepackage{amssymb}
\usepackage{amsthm}
\usepackage{amstext}
\usepackage{amsbsy}
\usepackage{amsopn}
\usepackage{amscd}
\usepackage{amsxtra}
\usepackage[colorlinks=true,allcolors=blue]{hyperref}
\usepackage{color}
\usepackage{soul}
\usepackage[dvipsnames]{xcolor}

\usepackage{ulem}

\def\phi{\varphi}
\def\epsilon{\varepsilon}

\newcommand{\setN}{\mathbb{N}}
\newcommand{\setR}{\mathbb{R}}

\newcommand{\setC}{\mathbb{C}}

\newcommand{\ket}[1]{| #1 \rangle }

\newcommand{\ii}{i}
\newcommand{\A}{\hat a}
\newcommand{\ad}{\hat a^\dagger}

\newcommand{\sigp}{\hat{\sigma}_+}
\newcommand{\sigm}{\hat{\sigma}_-}

\DeclareMathOperator*{\sign}{sign}

\everymath{\displaystyle}

\begin{document}

\title{Optimal control and selectivity of qubits in contact with a structured environment}





\author{Quentin Ansel}
\email{quentin.ansel@univ-fcomte.fr}
\affiliation{Institut UTINAM, CNRS UMR 6213, Universit\'{e} Bourgogne Franche-Comt\'{e}, Observatoire des Sciences de l'Univers THETA, 41 bis avenue de l'Observatoire, F-25010 Besan\c{c}on, France}

\author{Jonas Fischer}
\affiliation{Laboratoire Interdisciplinaire Carnot de Bourgogne, CNRS UMR 6303, Universit\'{e} Bourgogne Franche-Comt\'{e}, BP 47870, F-21078 Dijon, France}

\affiliation{Theoretische Physik, Universität Kassel, D-34132 Kassel, Germany}

\affiliation{Dahlem Center for Complex Quantum Systems and Fachbereich Physik, Freie Universität Berlin, Arnimallee 14, D-14195 Berlin, Germany}

\author{Dominique Sugny}
\affiliation{Laboratoire Interdisciplinaire Carnot de Bourgogne, CNRS UMR 6303, Universit\'{e} Bourgogne Franche-Comt\'{e}, BP 47870, F-21078 Dijon, France}

\author{Bruno Bellomo}
\affiliation{Institut UTINAM, CNRS UMR 6213, Universit\'{e} Bourgogne Franche-Comt\'{e}, Observatoire des Sciences de l'Univers THETA, 41 bis avenue de l'Observatoire, F-25010 Besan\c{c}on, France}


\begin{abstract}
We present a theoretical study of the optimal control of a qubit interacting with a structured environment. We consider a model system in which the bath is a bosonic reservoir at zero temperature and the qubit frequency is the only control parameter. Using optimal control techniques, we show the extent to which qubit population and relaxation effects can be manipulated. The reachable qubit states by a shaped control with a fixed maximum intensity are found numerically. We analyze the role of standard control mechanisms and the structure of the set of reachable states with respect to the coupling strength between the system and the environment. This investigation is used as a starting point to explore the selectivity problem of two uncoupled qubits interacting with their own baths and characterized by a specific coupling strength. We numerically derive the optimal control solution for a wide range of parameters and we show that the control law is close to a sinusoidal function with a specific frequency in some peculiar cases.
\end{abstract}

\maketitle

\section{Introduction}
Quantum optimal control~\cite{Boscain2021} is nowadays a key ingredient in a multitude of applications, extending from molecular physics~\cite{brifreview,stirapRMP,RMP19} to magnetic resonance and more recently to quantum technologies~\cite{glaserreview}. Efficient optimal control procedures have been developed to manipulate complex quantum systems toward various optimization targets~\cite{grape,gross,reichkrotov,alessandrobook,bonnardbook}.
  In spite of recent success, challenging control problems remain to be solved, in particular in the case of open quantum systems~\cite{koch:16}. The control of a dissipative system in the case of a non-structured environment~\cite{fixed,addis:2016} is by now well understood~\cite{glaserreview,altafini:04,koch:16,lapert:2010,mukherjee:2013,tannor:99,lapert:2013,lin1:20,lin:20,SH:11,stefanatos:09}. In addition, controllability results have been established in the mathematical literature for open quantum systems whose dynamics are governed by the Lindblad equation~\cite{altafini:03,dirr:09,omeara:12, Morzhin2021}. Control processes are not at the same stage of maturity for a structured bath~\cite{koch:16,Romano,schmidt2011,hall:2014,Arenz,
  RMPbreuer,RMPvega,fux2021} in a non-markovian regime when the fixed-dissipator assumption cannot be made~\cite{fixed,addis:2016}. Recently, several studies have focused on different aspects of the optimal control of non-Markovian dynamics. Among recent investigations, we mention the implementation of quantum gates~\cite{wilhelm:2009,tai:2014,hwang:2012,floether:2012}, quantum speed limit~\cite{deffner,mirkin:2016}, qubit purification~\cite{basilewitsch:2017,Fischer:19,Fischer:21}, thermalization~\cite{mukherjee:15}, generation of entanglement~\cite{wisniacki:19a,wisniacki:19b}, decoherence control~\cite{gordon,cui,mangaud}, the control of wave packet dynamics~\cite{ohtsuki}, controllability analysis~\cite{reich:15}, the control of an inhomogeneous spin ensemble~\cite{krimer}, and ground state cooling~\cite{triana}.

Despite these results, the importance of the information black-flow as a useful resource for control protocols remains an open issue. The information back-flow of an open quantum system is linked to the partial or total return of quantum excitations to the system after a certain amount of time spent in the environment. This effect is at the origin of non-Markovianity. We propose to strengthen the previous studies by analyzing the controllability and the reachable states of a qubit in contact with a structured environment. This analysis is a key prerequisite for determining control protocols performing specific tasks and using all the available resources of the environment. As an illustrative relevant example, we consider a qubit interacting with a Lorentzian bath (see~\cite{breuerbook,tai:2014} for general reviews on this system). A Markovian regime is achieved when the coupling strength and the detuning of the qubit with respect to the central bath frequency are small compared to the width of the Lorentzian spectral density. Non-Markovianity can be observed if these conditions are not met. The modulation of the qubit frequency allows us to modify the information back-flow, and to influence to some extent the relaxation effects. In this context, the role of a constant detuning is well known and the dynamics can be integrated analytically. Several studies have also pointed out the efficiency of a sinusoidal modulation with a specific frequency, usually called the magic frequency, as a way to prevent qubit relaxation~\cite{agarwal:99,janowicz:2000,macovei:14,franco:18}. However, a precise description of the reachable states for a fixed control time and of the corresponding control protocols is still lacking. We propose here to take a step in this direction by exploring numerically this control problem for the system under study. Note that a complete theoretical answer to this fundamental issue is a very difficult task and goes beyond the scope of this work. Using optimization procedures~\cite{glaserreview}, we find numerically the reachable states for a given initial configuration and we show the extent to which the qubit population can be manipulated in a fixed control time and a finite range of variations for the control parameter. We also discuss the underlying control mechanisms and the differences observed when the coupling strength is changed.

Such controllability results are the building block for control design. This general idea is exemplified in the case of the selective control of two qubits. Selectivity is an important prerequisite in quantum computing~\cite{glaserreview,vandamme:2018}, quantum sensing~\cite{calarco:20}, quantum discrimination and estimation~\cite{basilewitsch:20,audenarer:07}, and magnetic resonance~\cite{ma:2015,ansel:2017} applications. In particular, selective control is at the core of fingerprinting or contrast methods in magnetic resonance imaging or classical sensing approaches, which aim at exploiting the differences between qubit responses from a unique input excitation. Note that similar ideas have been also recently developed with the quantum Fisher information for quantum metrology applications~\cite{Lin2021,Pang2017}. On the basis of the controllability results, we investigate this control issue in the case of two qubits coupled to their own bath with different coupling strengths. Using numerical optimization, we derive the optimal control which brings one of the qubits to the ground state, while preventing the relaxation of the second qubit. We analyze the control mechanisms in different situations, as in the strong coupling regime, where the optimal control law is close to a sinusoidal function. We point out the properties of the system which are favorable to the selective control process.

The article is organized as follows. In Sec.~\ref{sec:model_syst}, the model system is presented with a specific attention to some limit cases. Section~\ref{sec:controllability} is dedicated to the numerical study of the reachable states by a control with a maximum intensity in a given control time. Different optimal control mechanisms are described and a comparison with standard control protocols is also carried out. In Sec.~\ref{sec:selectivity}, we study the selectivity problem of two qubits. Simple control solutions are derived and the physical limits of the process are found using numerical optimizations. Conclusion and prospective views are given in Sec.~\ref{sec:conclusion}. Technical details are reported in the Appendices~\ref{sec:Details on the model system}, \ref{sec:Details on the numerical simulations}, and \ref{sec:Ban-Bang control field}.
\section{The Model System}
\label{sec:model_syst}
We consider a qubit coupled to a bosonic reservoir at zero temperature, whose dynamics are governed by the Tavis-Cummings Hamiltonian~\cite{breuerbook}
\begin{equation}
\hat H(t) \!=\! \hbar \!\left(\!\omega_0(t) \sigp\sigm \!+\! \sum_l \!\left[ \omega_l \ad_l\A_l + g_l \sigp \A_l + g_l^* \sigm \ad_l \right] \!\right),
\label{eq:Hamiltonian}
\end{equation}
where $\omega_0(t) \in \setR$ is the qubit frequency, which can be controlled in time. The operators $\sigp$, $\sigm$ and $\ad_l$, $\A_l$ denote, respectively, qubit and cavity ladder operators. The parameters $\omega_l\in \setR$, and $g_l \in \setC$ are, respectively, the frequency of the mode $l$ and its coupling strength with the qubit. The bath is characterized by a Lorentzian spectral density~\cite{breuerbook} of the form
\begin{equation}
J(\omega) = \frac{\gamma}{2\pi}\frac{q^2}{(\omega-\omega_c)^2+q^2},
\end{equation}
where $q$ is the Lorentzian half width at half maximum. The parameter $\omega_c$ is the central frequency and $\gamma$ is an effective coupling strength. The bath correlation function $K$, which is connected to $J(\omega)$ by a Fourier transform, can then be written as
\begin{equation}
K(t-t')=\frac{\gamma q}{2}\exp [-q|t-t'|-i\omega_c(t-t')].
\label{eq:corr_fun_N=1}
\end{equation}

To simplify the following equations, we introduce the parameter $p=\sqrt{\frac{\gamma q}{2}}$ which has the dimension of a frequency. We assume that this model system is valid in the range of parameters we consider in this study: the couplings and the variations of $\omega_0$ must remain small with respect to $\omega_c$~\cite{agarwal}.

We investigate the case for which the whole system has a maximum of one excitation, so that the quantum state can be expressed as
\begin{equation}
\begin{split}
\ket{\psi(t)} = & c_0 \ket{\downarrow}_Q \otimes \ket{0}_B + c_1(t) \ket{\uparrow}_Q \otimes \ket{0}_B \\
& + \sum_l c_l(t) \ket{\downarrow}_Q \otimes \ket{l}_B,
\end{split}
\label{eq:definition_quantum_state}
\end{equation}
where $\ket{\downarrow}_Q$ and $\ket{\uparrow}_Q$ are the ground and excited states of the qubit, whilst $\ket{0}_B$ is the ground state of the reservoir, and $\ket{l}_B$ is the bosonic state with one excitation in the mode $l$ and the other modes in their ground state (we stress that $l$ cannot be taken equal to the symbols 0 and 1 to indicate these modes since the coefficients $c_0$ and $c_1$ are already used in other parts of the global state). By inserting Eq.~\eqref{eq:corr_fun_N=1} and Eq.~\eqref{eq:definition_quantum_state} into the Schr\"odinger equation $d_t \ket{\psi(t)} = - \ii \hbar^{-1} \hat H(t) \ket{\psi(t)}$, we arrive at the following dynamical equation
\begin{equation}
\label{eq:diff_eq_N=1}
 ~~\frac{d}{dt} \left( \begin{array}{c}
c_1(t) \\
y (t)
\end{array} \right) = \left( \begin{array}{cc}
- \ii \omega (t) & -p \\
 p & - q
\end{array} \right) \left( \begin{array}{c}
c_1 (t) \\
y (t)
\end{array} \right),
\end{equation}
where $y$ is defined by
\begin{equation}
y(t) = \int_0^t  dt' c_1(t')\left[p e^{-q|t-t'|}e^{ - \ii  \omega_c (t-t')} \right] + \ii \frac{1}{p} \sum_l c_l(0),
\label{eq:y(t)}
\end{equation}
and the frequency $\omega(t)$ is given by the detuning $\omega(t) = \omega_0(t) - \omega_c$. We remark that in Eq.~\eqref{eq:diff_eq_N=1}, $p$ plays the role of another effective coupling strength. The variable $y$ allows us to study in a compact way the whole system dynamics. Its value depends directly on the state of the bath. Similar approaches can be found in~\cite{garraway:97,janowicz:2000}. Technical details about the derivation of Eq.~\eqref{eq:diff_eq_N=1}, as well as a generalization to an arbitrary number of Lorentzian modes are given in Appendix~\ref{sec:Details on the model system}. Note that  $c_0$, defined in Eq.~\eqref{eq:definition_quantum_state}, is a constant of motion that does not impact the dynamics of the reduced system Eq.~\eqref{eq:diff_eq_N=1}. Moreover, the coherences of the qubit are given by $c_0^* c_1$ and $c_0 c_1^*$, and then the time evolution of their module is entirely determined by $|c_1|$. For simplicity, we set $c_0=0$.

In the limit when $\omega,p\ll q$, a simple approximated differential equation can be derived for the parameter $c_1$. When $q \rightarrow \infty$, we stress that the term $e^{-q|t-t'|}$ in Eq.\eqref{eq:y(t)} behaves like $\tfrac{2}{q} \delta (t-t')$, with $\delta$ the Dirac distribution. Assuming that the bath is initially empty, we obtain $y(t) \approx \tfrac{p}{q} c_1 (t)$, and thus we get $d_t c_1 (t) = \left(- \ii \omega(t) - \tfrac{p^2}{q} \right) c_1(t)$. In this limit, we observe that the control parameter $\omega(t)$ allows us to control the phase of $c_1$ but we cannot modify the population decay. We have to go beyond this approximation to obtain a noticeable modulation of the qubit relaxation.
Note that the boundary between Markovian and non-Markovian regimes is non-trivial because it depends on the amplitude of the information back-flow. The non-Markovian character is well-defined in the case of a free evolution \cite{breuerbook} and can be determined from a measure of non-Markovianity, such as the BLP measure~\cite{RMPbreuer}. Its definition and its role in the case of a controlled system are less clear and they are still a subject of theoretical studies~(see Ref.~\cite{franco:18} for an application with a sinusoidal control). In the rest of the paper, it will be sufficient to distinguish the weak and strong coupling regimes, respectively given by $q\gg2 p $ and $q\ll 2 p $, without discussing the possible non-Markovian behavior of the system, whose analysis is not crucial for our study of the control processes.

An interesting property of Eq.~\eqref{eq:diff_eq_N=1} is its linearity with respect to $c_1$ and $y$. This time-dependent differential system can thus be formally integrated by using the evolution operator
\begin{equation}
U(t) = \mathbb{T} \exp \left[ \int_0^t dt' \left( \begin{array}{cc}
- \ii \omega (t') & -p \\ p
 & - q
\end{array} \right) \right],
\label{eq:evolution_op_general}
\end{equation}
where $\mathbb{T}$ is the time-ordering operator. The system controllability can be deduced by writing $U$ as
\begin{equation}
\begin{split}
U(t) =  &\left( \begin{array}{cc}
e^{- \ii W (t)} & 0 \\
0 & e^{- q t}
\end{array} \right) \\
& \times \underbrace{\mathbb{T} \exp \left[ \int_0^t dt' \left( \begin{array}{cc}
0 & -p A(t') \\
p A^{-1}(t') & 0
\end{array} \right) \right]}_{h(W)},
\end{split}
\label{eq:evolution_operator_IR}
\end{equation}
with $W(t) = \int_0^t \omega (t') dt'$, and $A(t) = e^{-qt + \ii W(t)}$. The time-ordered exponential belongs to the group $SL(2,\setC)$ since the argument of the exponential function is a linear combination with complex coefficients of matrices having the same form of $\sigp$ and $\sigm$. However, due to the constraints on the coefficient $A(t)$, the set of admissible matrices is only a subset of $SL(2,\setC)$. Using Eq.~\eqref{eq:evolution_operator_IR}, it is straightforward to show that $\det (U(t)) = \exp[-qt - \ii W (t)]$, implying that the determinant of $U$ decreases as a function of time. We thus deduce that the system is not completely controllable since the identity operator cannot be generated for $t>0$ (see Ref.~\cite{sachkov:00}, theorem 2.9). This first analysis does not give a precise information about the reachable states by the qubit. This issue is adressed by means of numerical simulations in Sec.~\ref{sec:controllability}.
\section{Qubit controllability}
\label{sec:controllability}
This section first aims at describing the reachable states of the qubit by using numerical optimal control techniques. We analyze in a second step the corresponding control mechanisms and we review different state-of-the-art solutions.
\subsection{Reachable states of the qubit}
\begin{figure*}[t]
\begin{center}
\includegraphics[width=\textwidth]{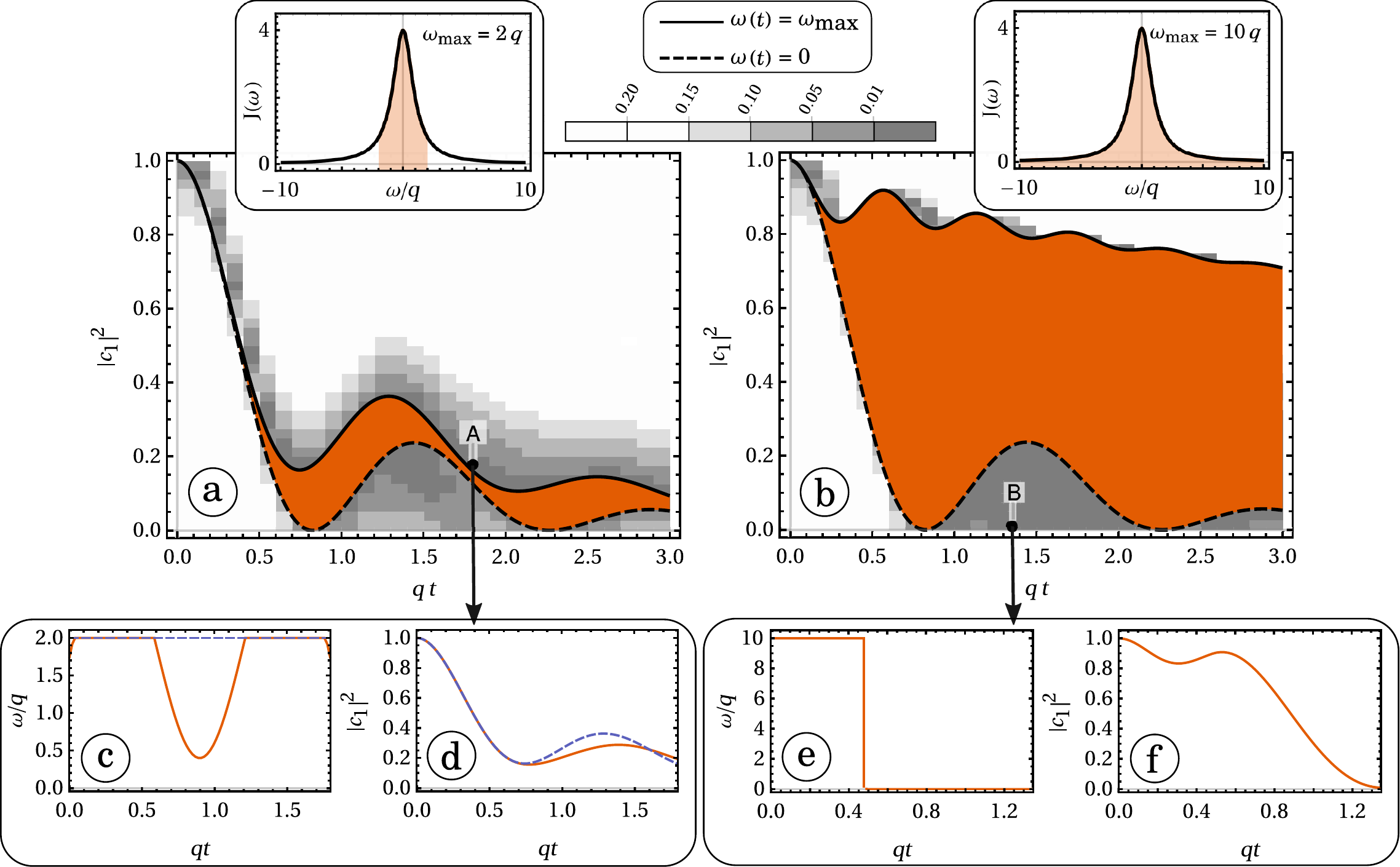}
\end{center}
\caption{Reachable states for two different bounds of the control amplitude: $\omega_{\text{max}}= 2q$ (a) and $\omega_{\text{max}}= 10q$ (b). A color code from white to dark gray gives the final value of $C_\diamond$ after optimization. Light gray areas correspond to states with a low cost $C_\diamond$, but not enough to ensure the reachability of the state (see the text  for details about the used criterion). The orange regions are reachable with a constant control field $\omega (t)\in [-\omega_{\text{max}},\omega_{\text{max}}]$. Solid and dashed black lines represent the trajectories with respectively $\omega(t) = \omega_{\text{max}}$ and $\omega (t)= 0$. The two insets at the top of the figure show the bath spectral density (black curves) while the colored interval depicts the range of possible frequencies for the control. The parameters are set to $p = \sqrt{5}q$ and, for the initial condition, $c_1(0) = 1$, $y(0)=0$. Panels (c), (d), (e), and (f) show examples of control fields and trajectories ending around the points A and B. In these panels, orange solid lines refer to the optimized solutions while in the panels (c) and (d) the dashed purple lines refer to the case of a constant control equal to $\omega_{\mathrm{max}}$.}
\label{fig:reachable_state_1}
\end{figure*}
We study the qubit controllability by computing the set of points $|c_1|^2(t)$ that can be reached in a fixed time $t$. For conciseness, we consider only the qubit population, and not its phase. Two relevant situations are used to illustrate the properties of the dynamics, namely the trajectories starting either from $\{c_1(0) = 1, y(0)=0\}$ (\textit{case 1}) or from $\{c_1(0) = 0, y(0)=1\}$ (\textit{case 2}). The the set of reachable states (reachable set) is computed as follows. We first partition the space of possible values of $|c_1|^2$ in the form $(t^\diamond,|c_1^\diamond|^2)$ where $\diamond$ denotes a point in the discretization grid. Using the algorithm GRAPE~\cite{grape}, we then search for a control field $\omega(t)$ connecting the initial state to the state $|c_1^\diamond|^2$ at time $t^\diamond$. For that purpose, we introduce the cost functional $C_\diamond= ||c_1(t^\diamond)|^2-|c_1^\diamond|^2|$, where $|c_1(t^\diamond)|^2$ is the population at time $t^\diamond$ generated by the optimal control field $\omega(t)$. The target state is said to be reached numerically if the final cost is lower than 0.01. Further details on the numerical simulations are reported in Appendix~\ref{sec:Details on the numerical simulations}.

\begin{figure}[h]
\begin{center}
\includegraphics[scale=0.7]{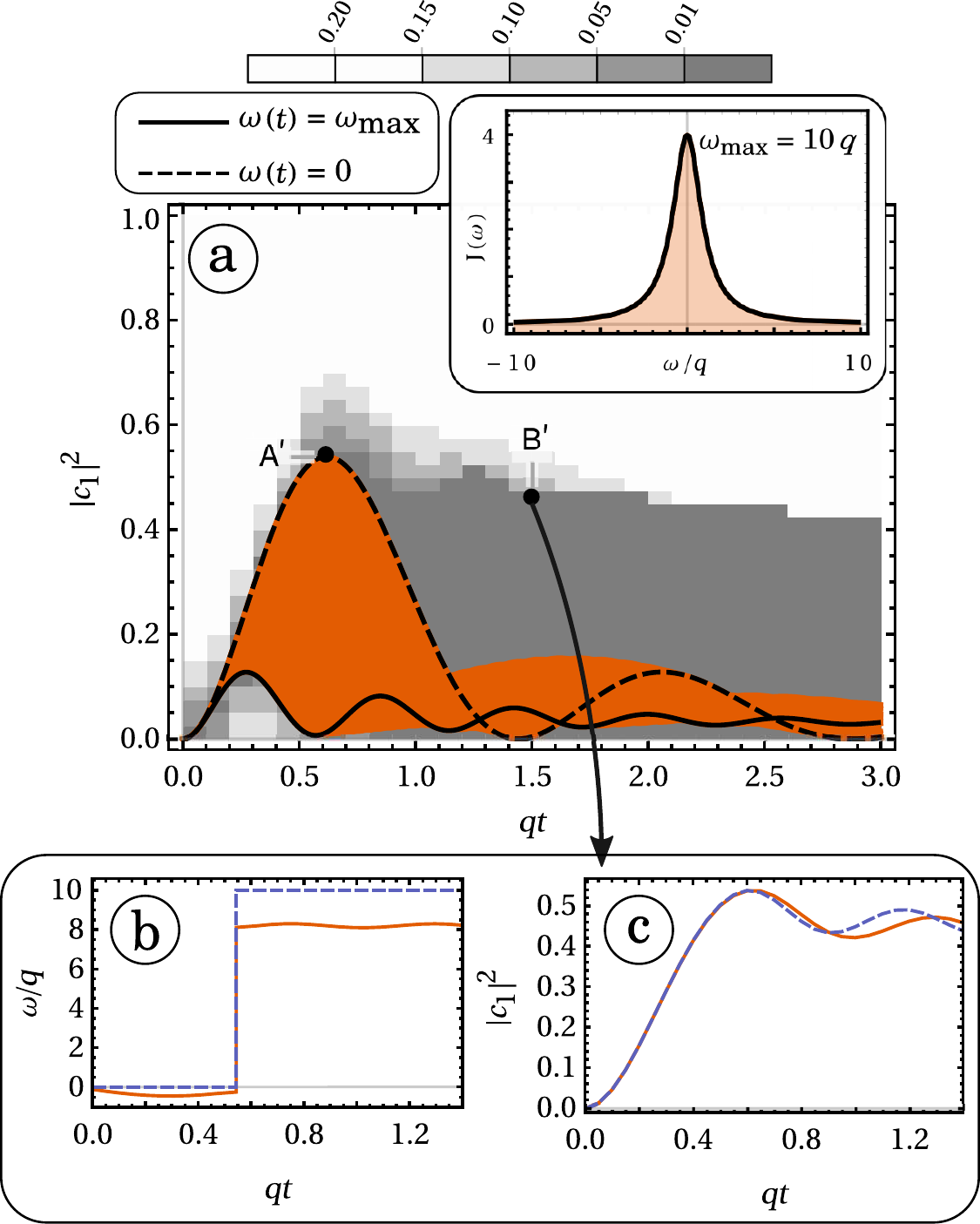}
\end{center}
\caption{Panel (a) is the same as Fig. \ref{fig:reachable_state_1} (with $p=\sqrt{5} q$), but with the initial conditions $c_1(0) = 0$, and $y(0)=1$. The maximum amplitude is set to $\omega_{\text{max}}=10 q$. Panels (b) and (c) show respectively two examples of control fields and the corresponding dynamics for a target state $|c_1|^2 \approx 0.45$ at $qt_f=1.4$ (point B$^\prime$). At this time, the maximum possible value of $|c_1|^2$ is $0.46$.  In these panels, orange solid lines refer to the optimized solutions while the dashed purple lines to a simpler piecewise-constant control.}
\label{fig:reachable_state_2}
\end{figure}

In order to limit the intensity of the control field, we introduce bounds on the control amplitude, $\omega (t)\in [-\omega_{\text{max}},\omega_{\text{max}}]$. Figures~\ref{fig:reachable_state_1} and \ref{fig:reachable_state_2} illustrate the different results. We discriminate in the two figures the points that can be reached by a constant or by a time-dependent control field. Figure~\ref{fig:reachable_state_1} shows that the relaxation effect is reduced when $\omega_{\text{max}}$ increases. Note that only a small increase of the reachable set is achieved with a time-dependent control over a constant control $\omega(t)=\omega_{\text{max}}$, $\forall t$, as can be seen in Figs.~\ref{fig:reachable_state_1}a, c, and d. On the other hand, shaped control fields allow us to access a large area of states below the trajectory with $\omega (t)= 0$. In this case, numerical simulations reveal that it is advantageous to use a piecewise-constant control, with first $\omega(t)=\omega_{\text{max}}$, and then $\omega(t)=0$, to reach states with zero population. Similar behaviors are observed in Fig.~\ref{fig:reachable_state_2} except that non-constant control fields allow us to explore a larger area of states that cannot be reached with constant fields. As can be observed in Figs.~\ref{fig:reachable_state_2}b and c, control protocols starting from $\omega \simeq 0$ and switching to $\omega \simeq \omega_{\text{max}}$ when $|c_1|^2(t)$ is maximum lead to $|c_1|^2$ around 0.5 even for long control times. High efficiency of the control process is achieved using small variations of control amplitude.

\begin{figure}[h]
\begin{center}
\includegraphics[scale=0.7]{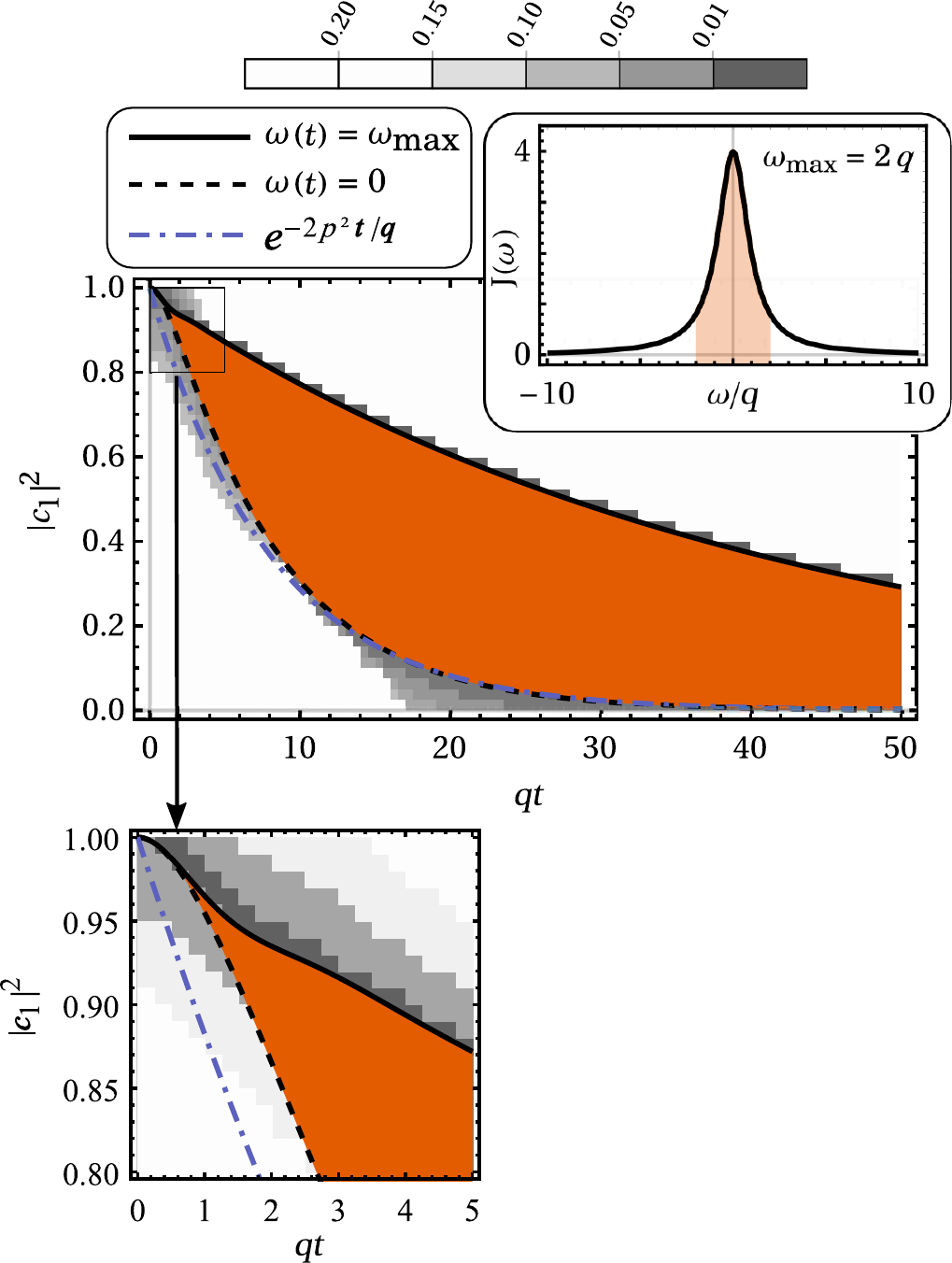}
\end{center}
\caption{Same as Fig.~\ref{fig:reachable_state_1}, but for a low coupling constant $p=0.25 q$, and $\omega_{\text{max}} = 2q$. The purple dashed-dotted curves correspond to $|c_1(t)|^2 = e^{-2p^2 t/q}$, which is the system trajectory in the limit $p,\omega_{\text{max}} \ll q$. Notice that due to the discretization, small gray areas can be seen above the curve corresponding to $\omega(t) = \omega_{\text{max}}$, $\forall t$, but if they are crossed by the black solid line, they only reflect the fact that the area can be reached by a constant field with a fidelity smaller than $0.01$ (see the zoomed inset in the bottom part of the figure).
}
\label{fig:reachable_state_3}
\end{figure}

The examples of Figs.~\ref{fig:reachable_state_1} and \ref{fig:reachable_state_2} are given for a qubit strongly coupled to the bath. In the weak coupling regime, the amplitude of oscillations is 
smaller, which leads to a much smaller reachable set. In particular, the gray area near $|c_1|^2 = 0$ disappears and the fastest way to steer the system to the ground state is approximately given by an exponential decay of the form $|c_1(t)|^2 = e^{-2p^2 t/q}$, which corresponds to the system trajectory in the limit $p,\omega_{\text{max}} \ll q$. This is illustrated in Fig.~\ref{fig:reachable_state_3}. In the weak coupling regime, almost all the accessible states can be reached by using constant controls.

\subsection{Control mechanisms}

The examples of control protocols given in Figs.~\ref{fig:reachable_state_1} and \ref{fig:reachable_state_2} suggest that most of the reachable states can be obtained with simple control fields, composed of one or two constant parts. In these cases, the dynamics can be solved analytically and simple control mechanisms can be highlighted. The evolution operator for a constant field $\omega$ during a time $t$ is given by
\begin{equation}
U (\omega,t)= \left(\begin{array}{cc}
U_{11}(\omega,t) &U_{12}(\omega,t) \\
-U_{12}(\omega,t) & U_{22}(\omega,t)
\end{array}  \right),
\end{equation}
where
\begin{equation}\label{eqBB}
\begin{split}
U_{11}(\omega,t) &= e^{-(q+\ii \omega)  t/2}\left[ \text{ch}\left(\frac{\Omega  t}{2} \right) + \frac{q-\ii \omega}{\Omega} \text{sh} \left(\frac{\Omega  t}{2} \right) \right], \\ \\
U_{22}(\omega,t) & = e^{-(q+\ii \omega) t/2}\left[ \text{ch}\left(\frac{\Omega  t}{2} \right) - \frac{q-\ii \omega}{\Omega} \text{sh} \left(\frac{\Omega  t}{2} \right) \right], \\\\
 U_{12}(\omega,t) & =-p\frac{2}{\Omega} e^{-(q + \ii \omega) t/2} \text{sh} \left(\frac{\Omega  t}{2} \right),
\end{split}
\end{equation}
and $\Omega= \sqrt{(q - \ii \omega)^2 - 4p^2}$. It is then straightforward to deduce the time evolution of $c_1(t)$ corresponding to a constant or to the concatenation of two constant fields. For instance, in the case of Fig.~\ref{fig:reachable_state_2}c, the time evolution of $c_1$ is given by $c_1(t)=U_{12}(0,t)$ if $t\leq t^\star$ and $c_1(t)=U_{11}(\omega_{\text{max}},t-t^\star)U_{12}(0,t^\star)+U_{12}(\omega_{\text{max}},t-t^\star)U_{22}(0,t^\star)$ if $t>t^\star$. The time $t^\star$ is the time when the field value is switched, here chosen as the position of the maximum of $|U_{12}(0,t)|^2$, given by the point A$^\prime$ in Fig.~\ref{fig:reachable_state_2}a. It is given by
\begin{equation}
\label{eq:max_abs_U12}
t^\star = \left[\frac{2}{\mathrm{Im} (\Omega)} \arccos \left( \frac{q}{\sqrt{q^2 + (\mathrm{Im} (\Omega))^2}}\right)\right]_{\omega = 0}.
\end{equation}

We can show that the modulus of $U_{12}$ is smaller than one, and therefore, it is not possible to completely transfer the bath excitation to the qubit. Moreover, we notice that $|U_{12}|$ is proportional to $ 1/|\Omega|$ and it decreases to zero when $\omega \rightarrow \pm \infty$. Thus, we recover the possibility to reduce the qubit relaxation when the qubit is detuned from the central bath frequency $\omega_c$.

\begin{figure}[t!]
\begin{center}
\includegraphics[width=\columnwidth]{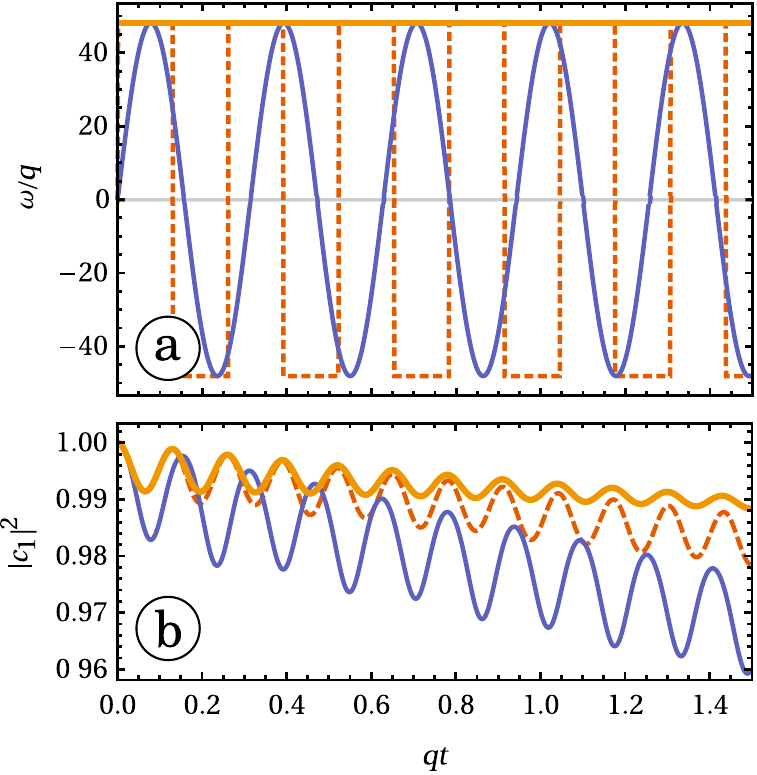}
\end{center}
\caption{Panel (a) displays the time evolution of the controls fields $\omega (t) = \omega_{\text{max}}$ (yellow thick solid line), $\omega (t) = \omega_{\text{max}} \sign(\sin(\omega_{\text{max}} t/2))$ (orange dashed line), and $\omega (t) = \omega_{\text{max}} \sin(\Theta t)$ (purple thin solid line). Panel (b) displays the population $|c_1|^2(t)$ associated with each control of panel (a). The parameters are set to $c_1(0)=1$, $y(0)=0$, $\Theta = 20 q$, $\omega_{\text{max}}=2.40483 \Theta $, and  $p = \sqrt{5}q$ (see the text for details).}
\label{fig:constant_vs_modulated}
\end{figure}

We conclude this section by a comparison between a sinusoidal control $\omega(t) = \omega_{\text{max}} \sin (\Theta t)$, proposed in Refs.~\cite{agarwal:99,janowicz:2000}, and two other simple control fields. It can be shown that a sinusoidal frequency modulation can strongly limit the qubit relaxation if $\omega_{\text{max}} \gg q,p $ (called below \textit{condition 1}), and additionally, if the oscillation frequency $\Theta$  is tuned to the so-called magic frequency defined by $J_0(\omega_{\text{max}}/\Theta)= 0$ (\textit{condition 2}), where $J_0$ is the first-order Bessel function \cite{janowicz:2000}. Further details about this control protocol are given in Appendix~\ref{sec:Ban-Bang control field}. We observe in Fig.~\ref{fig:constant_vs_modulated} that, in average, a sinusoidal control is less efficient than a constant control of amplitude $\omega_{\text{max}}$. This is easily explained by the fact that the effects of condition 2 on slowing down the decay hold only if condition 1 is verified. However, the time evolution of the population $|c_1|^2$ has a different period in the two cases, and maxima for a sinusoidal excitation can be slightly above the curve of a constant field. Then, a sinusoidal control at the magic frequency has here effects similar to those produced by the controls plotted in Fig.~\ref{fig:reachable_state_1}c and Fig.~\ref{fig:reachable_state_2}b. We have observed that the magic frequency mechanisms can be used with other periodic controls. A simple example is given by a square-wave control of period $4 \pi / \omega_{\text{max}}$, which defines another magic frequency limiting the qubit relaxation. The square-wave control field and the corresponding trajectory are plotted in Fig.~\ref{fig:constant_vs_modulated}. An intermediate decay with respect to the constant and the sinusoidal control cases is observed. Technical details about the square-wave control field are given in Appendix~\ref{sec:Ban-Bang control field}.

To summarize, we have numerically determined the reachable states of the qubit, showing the extent to which the qubit relaxation can be manipulated by the variation of the qubit frequency. The reachable sets with constant or time-dependent fields are very similar to the case of weak coupling, while significant areas can only be reached using a shaped control when the interaction between the system and the bath is strong. The reduction of relaxation effects by shaped controls is quite limited, but these controls can represent an efficient tool to reach quickly the ground state of the system. As shown in the different numerical simulations, this observation depends strongly on the characteristics of the environment. We study an application of these properties with the selectivity problem in Sec.~\ref{sec:selectivity}.
\section{Selectivity of two uncoupled qubits}
\label{sec:selectivity}
We consider the simultaneous control of two uncoupled qubits each interacting with its own bath, with two different coupling strengths, $p^{(1)}$ for the first qubit and $p^{(2)}=p^{(1)}(1+\alpha)$ for the second one, with $\alpha \in \setR$ a scale parameter. The value of $q$ is chosen to be equal for the two qubits. We assume that the two qubits are in the excited state at the initial time. The goal of the control is then to bring one of the qubits to the ground state at time $t_f$, while preventing the relaxation of the second qubit.

\begin{figure}[t]
\begin{center}
\includegraphics[width=\columnwidth]{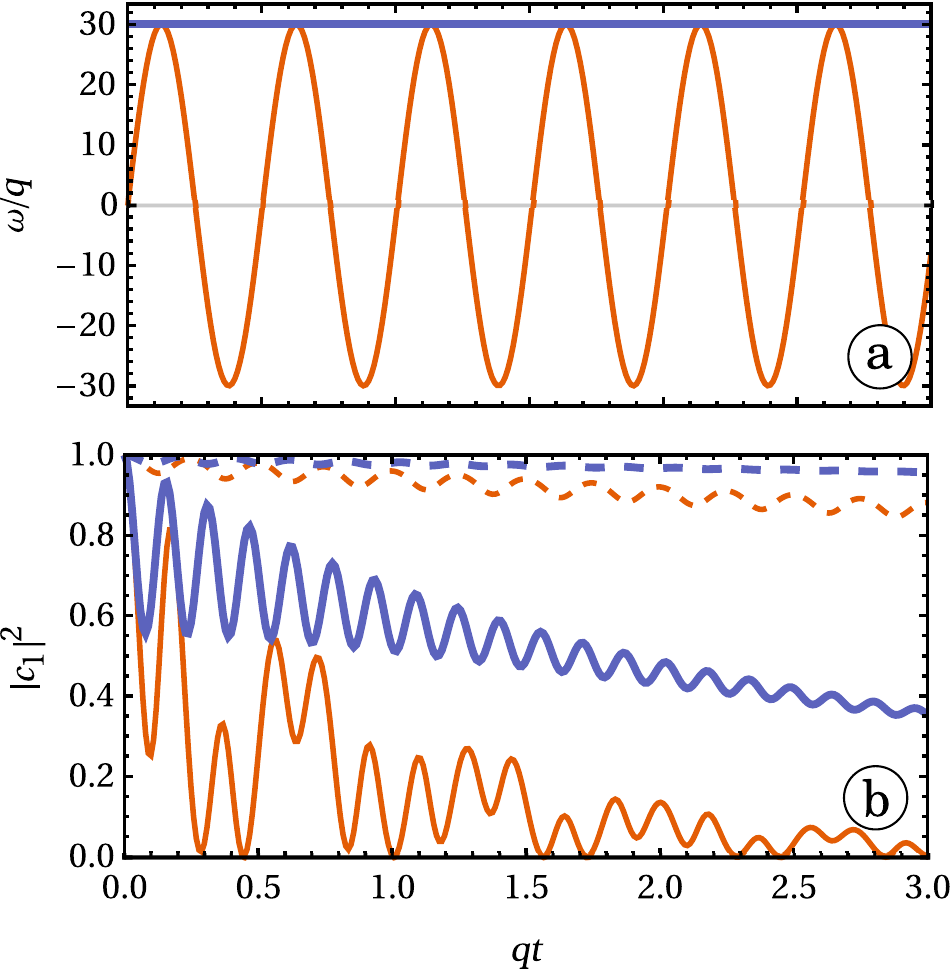}
\end{center}
\caption{Panel (a) displays the time evolution of constant (purple thick line) and sinusoidal (orange thin line) control fields used to discriminate qubits dynamics. Panel (b) depicts the evolution of qubits population. The same color and thickness of panel (a) are used. The two qubits have respectively a coupling strength $p^{(1)} = \sqrt{5}q$ (dashed line) and $p^{(2)}=\sqrt{5}(1+5)q$ (solid line). Other parameters are $c_1(0)=1$, $y(0)=0$, $\omega_{\text{max}}=30q$, and $\Theta = \omega_{\text{max}}/2.40483$.}
\label{fig:selectivity_sin_fun} 
\end{figure}
We first solve this control problem by using a simple frequency modulation in the case when the coupling strengths for the two qubits are quite different. Numerical simulations show that a sinusoidal control field at the magic frequency leads to an impressive gain of selectivity compared to a constant control at maximum amplitude. An example is plotted in Fig.~\ref{fig:selectivity_sin_fun}. An off-resonance effect (condition 1) prevents relaxation for the qubit with a small value of $p$, while a rapid loss of population is observed for the second qubit (which is characterized by a larger value of $p$), leading to a big population difference, which is well larger than for a constant control (at $t = t_f$). This dissimilarity between the two fields comes from the variations of the sinusoidal control. Since condition 1 is not verified for the second qubit, the time spent near $\sin(\Theta t) =0$ is not negligible, and the relaxation effect is important. The same kind of mechanism is observed in Figs.~\ref{fig:reachable_state_1}e and f. Note that the control strategy also works in the weak coupling regime, but the trajectories are very close to exponential decays, and the control time is very long.

The control field producing the best selectivity is in general difficult to determine, especially for smaller values of $\alpha$ and specific control times $t_f$. For this case, we use optimal control techniques to find optimal selective control fields. The control problem is defined through the minimization of the cost functional
\begin{equation}
C= \lambda \left|c_1^{(2)}(t_f)\right|^2 - \left|c_1^{(1)}(t_f)\right|^2,
\label{eq:cost_selectivity}
\end{equation}
where $|c_1^{(1)}(t_f)|^2$ and $|c_1^{(2)}(t_f)|^2$ are the excited-state populations of, respectively, the first and the second qubit, at the final time $t_f$, and  $\lambda$ is a parameter weighting $|c_1^{(2)}(t_f)|^2$. A value of $\lambda > 1$ forces the algorithm to converge toward a solution where the excited-state population of the second qubit is close to zero. The ideal value to obtain in order to achieve a perfect selectivity is then -1. Interesting results have been obtained for $\lambda =2$ and $t_f = 1.225/q$. All the numerical optimizations discussed in this section are performed using these two values. More elaborated cost functionals can also be used depending on the parameters to discriminate. In order to describe the efficiency of the optimal solution with respect to the case $\omega(t) = 0$, $\forall t$, we introduce the gain $G$ defined as
\begin{equation}
G= \frac{\left| |c_1^{(1)}(t_f)_{\omega=\omega_{\mathrm{opt}}}|^2 -|c_1^{(2)}(t_f)_{\omega=\omega_{\mathrm{opt}}}|^2\right|}
{\left| |c_1^{(1)}(t_f)_{\omega=0}|^2 -|c_1^{(2)}(t_f)_{\omega=0}|^2\right|},
\label{eq:gain_selectivity}
\end{equation}
where $\omega_{\mathrm{opt}}$ is the optimized control field. This function provides additional information about the selectivity process, and it may be easier to interpret than the cost functional $C$. However, when the denominator of $G$ is very small, note that a small variation of this latter may induce a large change of $G$. For this reason, we are not interested in a precise value of $G$ (we use the cost function $C$ for that purpose), but we are looking for a global tendency given by $G\gg 1$ or $G\ll 1$.
\begin{figure}[t]
\begin{center}
\includegraphics[width=\columnwidth]{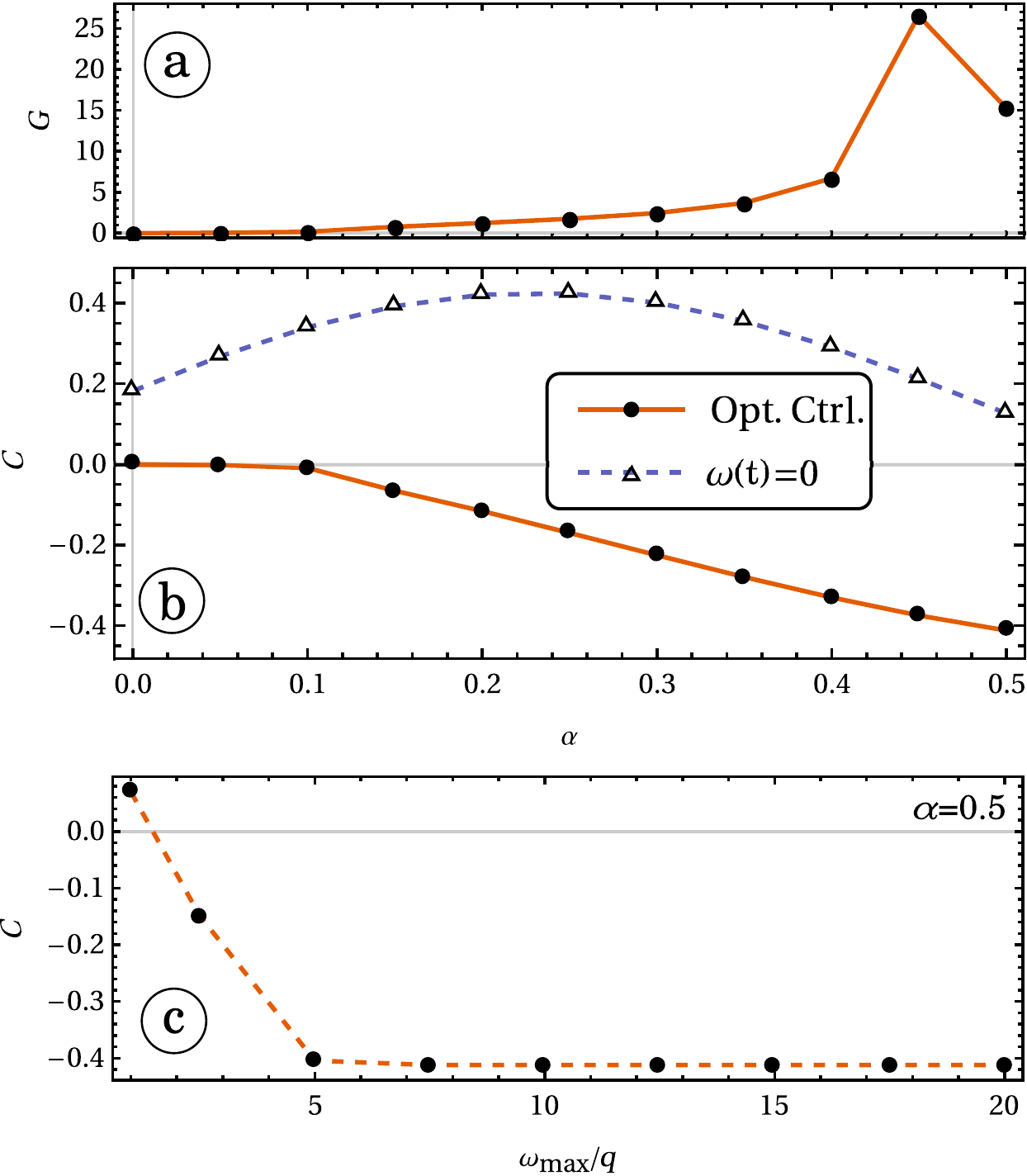}
\end{center}
\caption{Panels (a) and (b) show respectively the gain $G$ and the cost functional $C$ as a function of the scaling parameter $\alpha$, defined  by $p^{(2)} = p^{(1)}(1+\alpha)$ and $p^{(1)}=\sqrt{5}q$. These calculations are made without bounds on the control amplitude. Panel (c) displays the evolution of $C$ as a function of a bound on the control amplitude $\omega_{\text{max}}$, for  $p^{(1)}=\sqrt{5}q$ and $\alpha = 0.5$. For all the panels initial conditions are the same for the two qubits: $c_1(0) = 1$ and $y(0)=0$.}
\label{fig:selectivity}
\end{figure}
\begin{figure}[t]
\begin{center}
\includegraphics[width=\columnwidth]{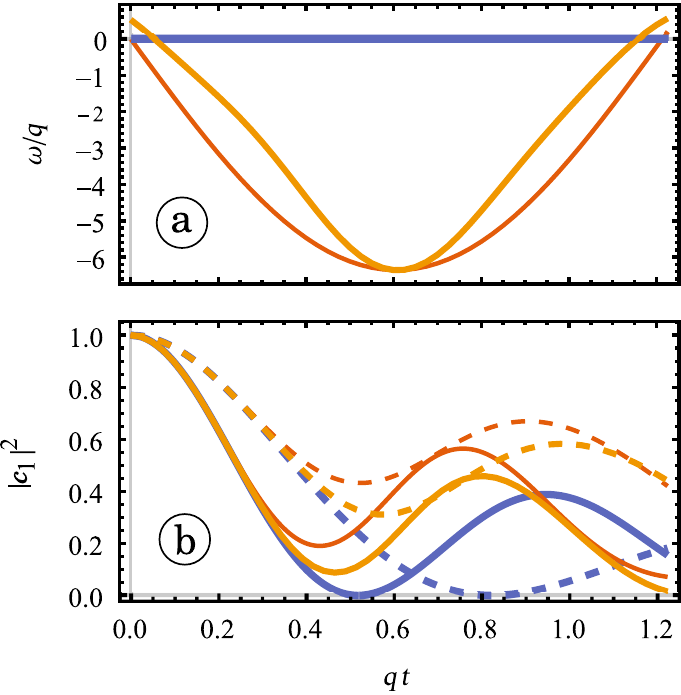}
\end{center}
\caption{Panel (a) represents the constant control $\omega(t) = 0$ (thick purple line), the sinusoidal control for $\omega_{\text{max}} = 6.36q$ and $\Theta = \omega_{\text{max}}/2.40483$ (orange thin line), and the optimized control (yellow medium line) for  $p^{(1)}=\sqrt{5}q$ and $\alpha = 0.5$. The time evolution of the population $|c_1|^2$ is plotted in panel (b). The first qubit is represented in dashed lines and the second one in solid lines. For all the panels initial conditions are the same for the two qubits: $c_1(0) = 1$ and $y(0)=0$.}
\label{fig:selectivity_opt_traj}
\end{figure}

The gain of selectivity and the cost functional for several values of $\alpha$ in the interval $[0,0.5]$ are plotted in Fig.~\ref{fig:selectivity}. We observe that the optimization process decreases significantly the cost functional. An impressive gain of selectivity is also achieved (up to 26 for $\alpha = 0.45$). Note that the cases $\alpha = 0.45$ and $\alpha = 0.5$ are very similar, and the large difference of $G$ is due to a small variation of the denominator in Eq.~\eqref{eq:gain_selectivity}. We also point out that the optimized control field not only enhances the population difference, but also forces the first qubit to be in the ground state at the final time. It is interesting to stress that the optimal solution does not require very large values of $\omega$, as illustrated in Fig.~\ref{fig:selectivity}c. The minimum $C \approx -0.412$ (for $\alpha = 0.5$) is reached at $\omega_{\text{max}} = 7.5 q$.

Figure~\ref{fig:selectivity_opt_traj} displays the optimal control field and the corresponding time evolution of the population $|c_1(t)|^2$ for $\alpha = 0.5$. A comparison is made between the optimal solution and the constant and sinusoidal cases. Extensive numerical simulations suggest that global minima have been reached. The same results are obtained with repeated launches of the algorithm GRAPE, initialized with multiple random initial fields (a similar study using a shooting algorithm~\cite{bonnardbook} has given us the same kind of results). The optimal field plotted in Fig.~\ref{fig:selectivity_opt_traj}, which is determined without control bounds has a maximum amplitude of $6.36 q$. Optimal control fields are very similar to sinusoidal functions (at the magic frequency) as can be seen in Fig.~\ref{fig:selectivity_opt_traj}. We can therefore conjecture that (slightly modified) sinusoidal frequency modulation plays a central role in selectivity mechanisms. These solutions could be an excellent starting point for more elaborated selective processes.

Finally, we point out that optimized control protocols are of limited interest in the weak coupling regime. In this regime, the population relaxation is very close to an exponential evolution. Using numerical observations, we notice that the qubit relaxation for a constant control ($\omega(t)=\omega_{\mathrm{max}}$, $\forall t$) is given by $|c_1(t)|^2 \simeq \exp \left(-2 p^2 q t /|q+\ii \omega_{\text{max}}|^2 \right)$, whilst for an arbitrary control field of maximum amplitude $\omega_{\mathrm{max}}$, we have $|c_1(t)|^2 \simeq \exp \left(-2 p^2 q f t /|q+\ii \omega_{\text{max}}|^2 \right)$. Here, $f$ is a parameter that depends non-trivially on the control field, but not on the parameter $p$. The effect of the control can be interpreted as a new time scale $t\rightarrow ft$, the optimization selecting a specific value of $f$. We deduce that the control process does not lead to a mechanism able to distinguish qubits dynamics. As an example, we consider a case similar to the ones described in Fig.~\ref{fig:selectivity_opt_traj}. We choose a sinusoidal control at the magic frequency, for $p^{(1)}=0.25q$, $\alpha=0.5$, and $\omega_\text{max} = 6.36q$. A fit with a very good accuracy of qubits trajectories is achieved for $f=3.38$. The ground state of the second qubit is reached approximately after a time $t_f$ given by $t_f = 5|q+\ii \omega_{\text{max}}|^2/( 2( p^{(2)})^2 q f) \approx 218/q$. We obtain a population of $|c_1^{(1)}(t_f)|^2 \approx 0.108$ and $|c_1^{(2)}(t_f)|^2 \approx 0.006$ for the first and the second qubit, respectively. The population difference is of the order of $0.10$ while it is estimated to $0.41$ in the strong coupling regime with the optimal solution given in Fig.~\ref{fig:selectivity_opt_traj}. Finally, note that, when both $\omega_\text{max}$ and $p$ are negligible with respect to $q$, we have $|q+\ii \omega_{\text{max}}|^2 \simeq q^2 $ and $f\approx 1$, and we recover the limit already discussed in Sec.~\ref{sec:model_syst} and \ref{sec:controllability}, such that $|c_1(t)|^2 \simeq \exp \left(-2 p^2 t /q^2 \right)$. In this second limit, the control protocol cannot have any efficient role in the selectivity process.

\section{Conclusion}
\label{sec:conclusion}

This paper presents a complete study of the optimal control of a qubit coupled to a structured environment and driven by modulation of its frequency. We have shown that the full system is not controllable and we have found numerically the reachable qubit states at a given time for two different initial configurations. We have described the structure of the reachable set when the interaction between the system and the environment is varied. We have observed that the structured environment offers interesting control opportunities in the strong coupling regime (for which the information back-flow may produce non-Markovian dynamics). In this case, different control mechanisms have been derived to steer the system to different targets. Differently, in the weak-coupling case, almost all the accessible states can be reached by constant controls. Using this preliminary study, we have explored the selective control of two qubits interacting with their own bath and with different coupling strengths. Optimal control computation leads to a control law close to a sinusoidal modulation. This result is based on two fundamental mechanisms which tend to enhance or prevent the relaxation effect by respectively tuning the qubit frequency at or out of resonance at specific times.

This study provides an example for the complete theoretical analysis of the control of an open quantum system using the tools of optimal control theory. We hope that this will inspire investigations in other quantum systems, and thus contribute to progress in quantum technologies that cannot be put forward without precise control of the different states involved in the dynamic processes. This work also opens the door to further studies on the selective control of qubits. The results of this paper are, e.g., a possible starting point for generalizing fingerprinting processes~\cite{ansel:2017} to the non-Markovian regime. Such protocols combined with optimal control techniques provide a way to approach the physical limits of a measurement process in terms of precision and sensitivity. It would be also interesting to study the link with the maximization of the quantum Fischer information in this kind of systems~\cite{Lin2021,Pang2017}.

\section*{Acknowledgments}

This work was mainly supported by the French ``Investissements d'Avenir'' program, project ISITE-BFC (Contract No.~ANR-15-IDEX-03). This research has been partially supported by the ANR projects COQS (Contract No.~ANR-15-CE30-0023-01). This project has received funding from the European Union's Horizon 2020 research and innovation programme under the Marie Skłodowska-Curie grant agreement No.~765267 (QuSCo).

\appendix

\section{Technical description of the system}
\label{sec:Details on the model system}
This appendix describes the different steps to derive Eq.~\eqref{eq:diff_eq_N=1}. We also generalize the model of Sec.~\ref{sec:model_syst} to the case when the bath can be characterized by an arbitrary number of Lorentzian modes.

We consider the interaction picture by using the unitary transformation $e^{-\ii t \sum_l \omega_l \ad_l\A_l }$. The Hamiltonian defined in Eq.~\eqref{eq:Hamiltonian} transforms into
\begin{equation}
\hat H_I(t) \!=\! \hbar \!\left(\!\!\omega_0(t) \sigp\sigm \!+\! \!\sum_l \!\left[ g_l e^{-\ii \omega _l t }\sigp \A_l\! +\! g_l^* e^{\ii \omega _l t}\sigm \ad_l \right]\! \!\right)\!,
\label{eq:full_Hamiltoanian}
\end{equation}
and the corresponding Schr\"odinger equation is
\begin{equation}
\begin{split}
d_t\ket{\psi(t)}\! = \!&\underbrace{-\ii \left(\! \omega_0(t) c_1(t)+\! \sum _ l g_l e^{-\ii \omega_l t }c_l(t)\right)}_{d_t c_1(t) \!} \! \ket{\uparrow}_Q \otimes \ket{0}_B  \\
&\! \underbrace{- \ii \sum_l g_l^* e^{\ii \omega_l t} c_1 (t)}_{d_t c_l (t)}\!\ket{\downarrow}_Q \otimes \ket{l}_B.
\end{split}
\label{eq:schrodinger_eq}
\end{equation}
Integrating formally $d_t c_l(t)$, we get
\begin{equation}\label{eqc1}
c_l (t) = c_l(0) - \ii \int_0^t dt' g_l^* e^{\ii \omega_l t'} c_1 (t').
\end{equation}
Plugging Eq.~\eqref{eqc1} into $d_t c_1(t)$ of Eq.~\eqref{eq:schrodinger_eq} leads to
\begin{equation}
\begin{split}
d_t c_1(t) = &- \ii \omega_0 (t) c_1(t)  \\
& -  \int_0^t  dt' c_1(t') \underbrace{\sum_l |g_l|^2 e^{-\ii \omega_l (t-t')} }_{K(t-t')} - \ii \sum_l c_l(0).
\end{split}
\label{eq:diff_eq_c1}
\end{equation}
To proceed further, a bath correlation function $K(t-t')$ has to be specified. We assume that $K$ is the Fourier transform of a spectral density distribution $J$ given by a sum of Lorentzians centered around the frequency $\omega_c$~\cite{breuerbook}. We then have
\begin{equation}
K(t-t') = \sum_{k=1}^N p_k^2 e^{-q_k|t-t'|}e^{ - \ii  \omega_c (t-t')} ~ ; ~ p_k^2, q_k \in \setR^+.
\label{eq:corr_fun}
\end{equation}
Inserting Eq.~\eqref{eq:corr_fun} into Eq.~\eqref{eq:diff_eq_c1} gives
\begin{equation}
d_t c_1(t) = - \ii \omega_0 (t) c_1(t) - \sum_{k=1}^N p_k y_k (t),
\label{eq:d_t c_1(t) }
\end{equation}
where
\begin{equation}
y_k(t)\! =\! \int_0^t \! dt' c_1(t')\!\left[p_k e^{-q_k|t-t'|}e^{ - \ii  \omega_c (t-t')} \right] +\! \ii \frac{1}{p_k}\!\sum_l c_l(0).
\end{equation}
Since $|t-t'|\geq 0$, we can differentiate $y_k$:
\begin{equation}
d_t y_k(t) = -(q_k + \ii \omega_c) y_k(t) + p_k c_1 (t).
\label{eq:d_t y_k(t) }
\end{equation}
We observe that the whole dynamics is given by a system of $N+1$ first order linear differential equations. The first coordinate $c_1$ encodes the qubit dynamics ($c_0$ is a constant), while the $y_k$ variables are associated with the state of each effective mode. In the main text, we consider only the case $N=1$. To simplify the notation we use $y_1 = y$, and the system of differential equations is rewritten in a matrix form, to obtain Eq.~\eqref{eq:diff_eq_N=1}.

\section{Details on the numerical simulations}
\label{sec:Details on the numerical simulations}

 The numerical calculations discussed in this paper have been performed with piecewise-constant control fields. The time step is 0.02/$q$ except in the case of the long control sequences of Fig.~\ref{fig:reachable_state_3} where it is  0.1/$q$. The number of points in a grid  $(t^\diamond,|c_1^\diamond|^2)$ is between 1271 and 4141. For each control field, 800 iterations have been used to obtain the convergence of the GRAPE algorithm. The computation time for a field is between 0.12 seconds (shortest values of $t^\diamond$) and 356 seconds (longest values of $t^\diamond$). These times are given for a single core clocked at 3.2GHZ. Since all optimizations are independent, parallel computing has been used to reduce the overall computation time. To speed up the calculations, in some cases we verified before the optimization process which partitions of the grid can be reached by constant controls.

\section{Sinusoidal and square-wave control fields}
\label{sec:Ban-Bang control field}
In this appendix, we present the construction of sinusoidal and periodic square-wave control fields of amplitude $\omega_{\text{max}}$ which lead to an effective decoupling between the qubit and the reservoir. First, we recall the main results that can be established in the case of sinusoidal control~\cite{agarwal:99,janowicz:2000}, and then we extend this approach to square-wave controls.

\subsection{Sinusoidal control}

A possible starting point is to consider the evolution operator given in Eq.~\eqref{eq:evolution_operator_IR}. This propagator is expressed as the product of two operators, one being diagonal, and the other having non-diagonal elements in the canonical basis. This latter is a function of a time-dependent coupling term which depends on the integral over time of the control. Explicitly, the effective coupling is proportional to $\exp\left(-\ii \int_0^t \omega(t')dt'\right)$. We consider a sinusoidal control field $\omega(t) = \omega_{\text{max}} \sin (\Theta t)$, for which an explicit calculation of the integral is possible. The effective coupling can be expressed as
\begin{equation}
\begin{split}
& e^{\ii \frac{\omega_{\text{max}}}{\Theta} \cos(\Theta t)} = \\
& J_0 \left(\frac{\omega_{\text{max}}}{\Theta} \right) + 2 \sum_{n=1}^\infty (\ii)^n J_n \left(\frac{\omega_{\text{max}}}{\Theta} \right) \cos (n \Theta t),
\end{split}
\label{eq:expansion_bessel}
\end{equation}
with $J_n$ the $n$-order Bessel function. A simple way to reduce the coupling between the system and the environment is to cancel the zeroth order term of the expansion, i.e., to impose $J_0 \left(\omega_{\text{max}}/\Theta \right) = 0$. The first zero of the function is given by $\omega_{\text{max}}/\Theta \approx 2.40483$. The corresponding solution is called the magic frequency. However, the effect of this choice on the dynamics is noticeable only if $\omega_{\text{max}}\gg q,p$. Otherwise, the system stays during a long time near resonance, and the relaxation is fast. It results that the control effect is very weak.

\subsection{Square-wave control}
We consider a control of period $T$ with values $+\omega_{\text{max}}$ and $-\omega_{\text{max}}$ during the first and second half periods, respectively. Assuming that the initial condition is $c_1(0)=1$ and $y(0)=0$, a straightforward calculation gives
\begin{equation}
\begin{split}
c_1(T) = & U_{11}(-\omega_{\text{max}},T/2)U_{11}(\omega_{\text{max}},T/2) \\
& - U_{12}(-\omega_{\text{max}},T/2) U_{12}(\omega_{\text{max}},T/2).
\end{split}
\label{eq:U11_bang_bang}
\end{equation}
In the limit $\omega_{\text{max}} \rightarrow \pm \infty$, we have $\Omega \simeq \ii(q \mp \ii \omega_{\text{max}})$, and then $U_{11} \rightarrow 1$ [see Eq.~\eqref{eqBB}]. We deduce that the dynamics is governed by the term $U_{12}$. Equation~\eqref{eq:U11_bang_bang} can be approximated as
\begin{equation}
\label{eqappr}
c_1(T) \simeq 1 - 2\frac{p^2}{|\Omega| ^2} e^{-qT/2}\left[ \cosh(qT/2)- \cos (\omega_{\text{max}} T/2) \right].
\end{equation}
Equation~\eqref{eqappr} shows that the relaxation effect is minimum when $\omega_{\text{max}} T/2 = 2 \pi$, since the first order term of a Taylor expansion in $T$ is zero. This constraint is the analog of $J_0(\omega_{\text{max}}/\Theta)=0$ for the sinusoidal control. On the opposite, the relaxation effect is enhanced if $\omega_{\text{max}} T/2 = (2k +1)\pi$, $k \in \setN$. It must be emphasized that the square-wave solution is always less efficient than a constant control of amplitude $\pm \omega_{\text{max}}$ and duration $T$ because in this case
\begin{equation}
c_1(T) = U_{11}(\pm \omega_{\text{max}},T),
\end{equation}
which is equal to $1$ in the limit $\omega_{\text{max}} \rightarrow \pm \infty$. The convergence toward 1 is therefore faster. An example of dynamics with a square-wave control is plotted in Fig.~\ref{fig:constant_vs_modulated}.


\bibliographystyle{apsrev4-1}

\end{document}